\documentclass[prl,twocolumn,superscriptaddress,noshowpacs]{revtex4}
\usepackage{graphicx}
\usepackage{physics}
\usepackage{bm}
\usepackage[utf8]{inputenc}

\begin{document}
\title{Quantum metric and wavepackets at exceptional points in non-Hermitian systems}

\author{D.~D.~Solnyshkov}
\affiliation{Institut Pascal, PHOTON-N2, Universit\'e Clermont Auvergne, CNRS, SIGMA Clermont, F-63000 Clermont-Ferrand, France.}
\affiliation{Institut Universitaire de France (IUF), F-75231 Paris, France}

\author{C.~Leblanc}
\affiliation{Institut Pascal, PHOTON-N2, Universit\'e Clermont Auvergne, CNRS, SIGMA Clermont, F-63000 Clermont-Ferrand, France.}

\author{L.~Bessonart}
\affiliation{Institut Pascal, PHOTON-N2, Universit\'e Clermont Auvergne, CNRS, SIGMA Clermont, F-63000 Clermont-Ferrand, France.}

\author{A. Nalitov}
\affiliation{Institut Pascal, PHOTON-N2, Universit\'e Clermont Auvergne, CNRS, SIGMA Clermont, F-63000 Clermont-Ferrand, France.}
\affiliation{Faculty of Science and Engineering, University of Wolverhampton, Wulfruna St, Wolverhampton WV1 1LY, UK}

\author{J. Ren}
\affiliation{Beijing Key Laboratory for Optical Materials and Photonic Devices, Department of Chemistry, Capital Normal University, Beijing 100048, People's Republic of China}
\affiliation{Tianjin Key Laboratory of Molecular Optoelectronic Sciences, Department of Chemistry, School of Sciences, Tianjin University, Collaborative Innovation Center of Chemical Science and Engineering, Tianjin 300072, People's Republic of China}

\author{Q. Liao}
\affiliation{Beijing Key Laboratory for Optical Materials and Photonic Devices, Department of Chemistry, Capital Normal University, Beijing 100048, People's Republic of China}

\author{F. Li}
\affiliation{Key Laboratory for Physical Electronics and Devices of the Ministry of Education \& Shaanxi Key Lab of Information Photonic Technique, School of Electronic and Information Engineering, Xi'an Jiaotong University, Xi'an 710049, China}

\author{G.~Malpuech}
\affiliation{Institut Pascal, PHOTON-N2, Universit\'e Clermont Auvergne, CNRS, SIGMA Clermont, F-63000 Clermont-Ferrand, France.}

\begin{abstract}
The usual concepts of topological physics, such as the Berry curvature, cannot be applied directly to non-Hermitian systems. 
We show that another object, the quantum metric, which often plays a secondary role in Hermitian systems, becomes a crucial quantity near exceptional points in non-Hermitian systems, where it diverges in a way that fully controls the description of wavepacket trajectories.
The quantum metric behaviour is responsible for a constant acceleration with a fixed direction, and  for a non-vanishing constant velocity with a controllable direction. Both contributions are independent of the wavepacket size.
\end{abstract}
\maketitle

\maketitle

Non-Hermitian systems in quantum mechanics \cite{Moiseyev2011} and optics \cite{Ozdemir2019} expand the spectrum of possibilities much beyond that of the Hermitian ones \cite{Konotop2016,Ganainy2018}. A topic of very strong interest interesting is offered by the merging of the so-called topological and non-Hermitian physics. For example, the non-Hermitian systems cannot be described by the same topological invariants as the Hermitian ones: it is not the Chern number, but the winding number of a complex effective field, which determines the topology in the non-Hermitian case \cite{Bliokh2017}.
Partially because of this, the quantum geometry of the eigenstates (determining the Chern number) has not been studied extensively in non-Hermitian systems  \cite{PhysRevA.86.064104,PhysRevA.99.032121,Brody2013}. This geometry is described by the quantum geometric tensor \cite{provost1980riemannian} (QGT), which includes the Berry curvature (the cornerstone of Hermitian topological physics) and the quantum metric. Even in Hermitian systems, the whole QGT has been  measured experimentally relatively recently \cite{gianfrate2020measurement,Yu2019} and the quantum metric often plays a secondary role with respect to the ubiquitous Berry curvature.
It is so far used in the calculations of quantum phase transitions \cite{Zanardi2007}, electronic orbital magnetic susceptibility \cite{Gao2014,Piechon2016}, excitonic levels \cite{PhysRevLett.115.166802}, and superfluidity in flat bands \cite{peotta2015superfluidity,Liang2017}. The anomalous Hall effect, corresponding to a lateral shift of a wavepacket (WP) evolving adiabatically within a single dispersion branch, involves both components of the QGT, but the dominant role is played by the Berry curvature, while the quantum metric appears as a correction \cite{Gao2014,Bleu2018effective}.  The situation is different in vicinity of the exceptional points in non-Hermitian systems. Such points arise where the Hermitian and non-Hermitian parts of the Hamiltonian exactly compensate each other. The exceptional points can be encircled either statically (in so-called "stroboscopic" experiments) or dynamically, by changing the Hamiltonian parameters over time. In the first case, one has the access to all branches of the complex energy dispersion, and the topological winding numbers \cite{Bliokh2017,Shen2018}  can be studied \cite{zhong2018winding}.  In the second case, the non-Hermitian properties of the system bring chirality into the dynamics of the system, making the exponentially decaying eigenstates completely inaccessible \cite{Berry2011a,Berry2011b}. In particular, it is actually impossible to encircle the exceptional point twice in any direction \cite{Milburn2015}. %A single circle with branch change can be adiabatic only in one direction. 
While from the Hermitian point of view, this may seem as a flaw, one can actually take advantage of this feature to construct chiral optical transmitters \cite{Doppler2016,Zhang2018}. The non-Hermitian systems thus require switching to a new way of thinking. Thus, very recent studies started to focus on the divergence of the quantum metric \cite{Brody2013,Berry2020}.

Many different experimental implementations of exceptional points are currently studied in photonics \cite{Ozdemir2019}. They include coupled waveguides and resonators, as well as various lattices \cite{Hahn2016,Zhang2016}, with foreseen applications such as enhanced sensors \cite{Wiersig2016} and quantum information \cite{Naghiloo2019}. Microcavities \cite{Microcavities}, with their widely tunable properties, represent a particularly versatile platform for non-Hermitian physics \cite{Gao2015b,Gao2018}. The singular optical axes, also called Voigt points \cite{Voigt1902}, represent a particularly interesting configuration, fascinating scientists for more than a century \cite{Berry2003}. They appear in optical systems combining spin-orbit coupling and polarization-dependent absorption \cite{Landau8,Sturm2020}. Recently, the square root topology of such points has been demonstrated in a ZnO-based microcavity \cite{PhysRevA.95.023836,Richter2019}. 
Contrary to other non-Hermitian systems, many of which are described by a synthetic parameter space (e.g. coupling constant, detuning, and gain/loss for coupled resonators), these points occur in the reciprocal space describing the direction of the propagation of a beam. This allows to study effects linked with the WP dynamics in real and reciprocal spaces, like the anomalous Hall effect in Hermitian systems. However, because of the gain and decay, we can expect the correct description to be essentially non-adiabatic, representing a superposition of branches, and thus not controlled by the Berry curvature any more.

In this work, we study the dynamics of WPs centered at an exceptional point, taking the Voigt point as a particular example. We demonstrate that the quantum metric of the eigenstates in the vicinity of exceptional points plays a dominant role in the WP dynamics, leading to a non-vanishing polarization-dependent group velocity.

We begin with the definition of the quantum metric, allowing to calculate the distances between the quantum states \cite{provost1980riemannian}. In general, such distance $ds$ between the states $\psi(\lambda)$ and $\psi(\lambda+\delta\lambda)$ is linked with their overlap:
\begin{equation}
\label{metric}
ds^2=g_{ij}d\lambda_id\lambda_j=1-|\bra{\psi(\bm{\lambda})}\ket{\psi(\bm{\lambda}+\delta\bm{\lambda})}|^2
\end{equation}
where the metric tensor $g_{ij}$ can be found as a real part of the QGT, $g_{ij}=\Re T_{ij}$:
\begin{equation}
    T_{ij}=\bra{\frac{\partial \psi}{\partial \lambda_i}}\ket{\frac{\partial\psi}{\partial\lambda_j}}-\bra{\psi}\ket{\frac{\partial\psi}{\partial\lambda_i}}\bra{\frac{\partial\psi}{\partial\lambda_j}}\ket{\psi}
    \label{qgt}
\end{equation}
The overlap integral between an arbitrary state $\ket{\psi}$ and another state $\ket{\psi_1}$ can therefore be written using the metric tensor $g_{ij}$ as
\begin{equation}
    I=1-\left(\int\limits_{\ket{\psi}}^{\ket{\psi_1}}\sqrt{g_{ij}d\lambda_i d\lambda_j}\right)^2
    \label{ovlint}
\end{equation}
where the integral should be taken along a geodesic line. These integrals determine the behavior of the coefficients $c_l(\bm{\lambda})$ ($|c_l|^2=I$) which define the representation of an arbitrary initial state $\psi$ as a superposition of the branches $l$ of the eigenstates $\ket{\psi_l(\bm{\lambda})}$ used in the general solution of the Schr\"odinger equation
\begin{equation}
    \ket{\psi(\bm{\lambda},t)}=\sum_l c_l(\bm{\lambda})e^{-i\frac{E_l(\bm{\lambda)}}{\hbar}t}\ket{\psi_l(\bm{\lambda})}
\end{equation}

\begin{figure}[tbp]
 \includegraphics[width=1.0\linewidth]{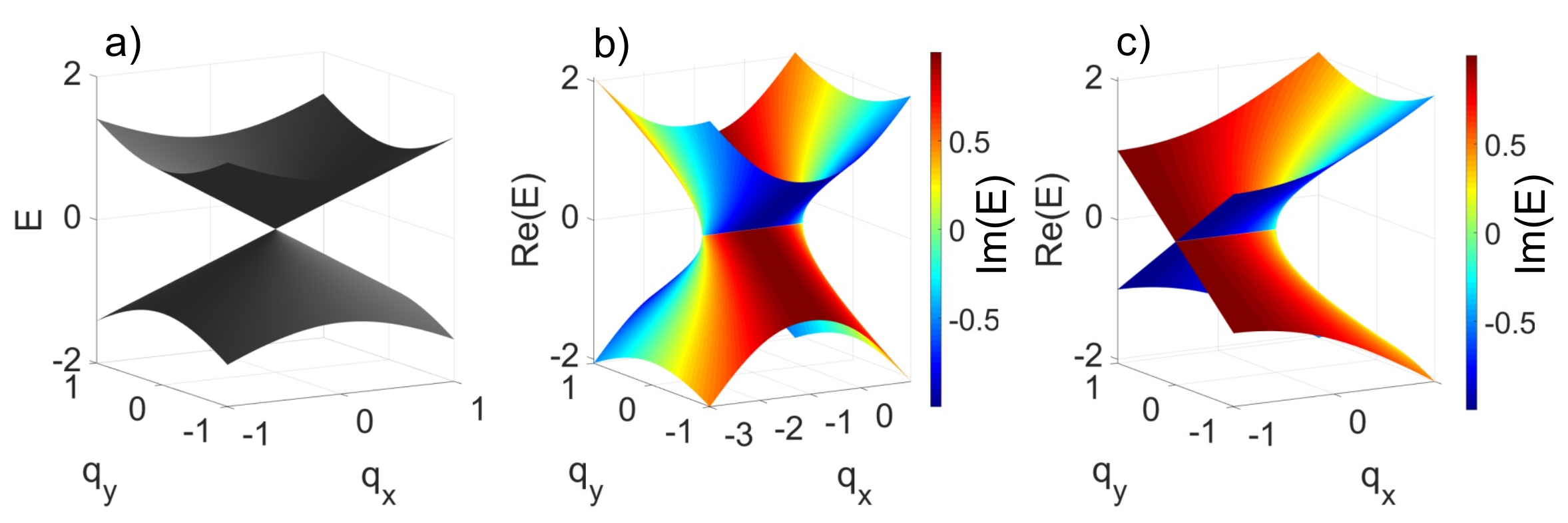}
 \caption{\label{fig1} a) Dispersion of a diabolical point (conical intersection). b) Transformation of a diabolical point into two exceptional points.  c) Zoom on a single exceptional point at the origin. The false color shows the imaginary part of the energy. }
 \end{figure}

\emph{Diabolical and exceptional points} We first consider a simple Hamiltonian with a diabolical point, which will serve as a reference: 
\begin{equation}
{\hat H_0} = \hbar c\left( {\begin{array}{*{20}{c}}
0&{q{e^{ - i\varphi }}}\\
{q{e^{i\varphi }}}&0
\end{array}} \right) = \hbar c\bm{q} \cdot \bm{\sigma}
\end{equation}
Here, $\bm{q}$ is a wave vector in polar coordinates ($\varphi$ is its polar angle), $c$ is the celerity, and $\bm{\sigma}$ is a vector of Pauli matrices.
This 2D massless Dirac Hamiltonian, written for light on the circular basis, actually describes the oldest known diabolical point, dating back to Hamilton \cite{Hamilton1837} and Lloyd \cite{Lloyd1837}.  The most striking manifestation of the diabolical point is the conical refraction, where a Gaussian WP is transformed into a ring. This is illustrated by Fig.~\ref{fig1}(a) showing the conical intersection at the diabolical point. In microcavities, such diabolical points can arise from the combination of TE-TM splitting and birefringence \cite{Tercas2014,gianfrate2020measurement}.

The dispersion is strongly modified in the presence of non-Hermitian terms in the Hamiltonian. Such terms can split a single diabolical point into two exceptional points (Fig.~\ref{fig1}(b)). If these points are sufficiently far in the parameter space, it becomes possible to study each of them separately (Fig.~\ref{fig1}(c)). We therefore consider a simplest Hamiltonian with an exceptional point in the center of the parameter space (reciprocal space), with $\hbar=1$:
\begin{equation}
\hat H_1 = \left( {\begin{array}{*{20}{c}}
0&{\alpha q{e^{ - i\varphi }} +a}\\
{\alpha q{e^{i\varphi }} + a}&0
\end{array}} \right) + i\left( {\begin{array}{*{20}{c}}
0&{-ia}\\
{ia}&0
\end{array}} \right)
\label{hamNH}
\end{equation}
The first part of the Hamiltonian describes a real $q$-dependent Rashba-type effective field $\alpha q$ and a constant contribution $a$ along $x$. The second part is a non-Hermitian imaginary part which is proportional to $\sigma_y$ and can be described as an imaginary effective field pointing along $y$. Its magnitude $a$ exactly equal to the constant real field along $x$. This configuration corresponds to the combination of TE-TM field, birefringence, and linear dichroism (polarization-dependent absorption) in the microcavities \cite{Richter2019}. Equivalent Hamiltonians have been considered in other works~\cite{Bliokh2017}.

The energy dispersion of this Hamiltonian is indeed very different from the case of the Dirac Hamiltonian: for small $q$, $E(q)=\pm\sqrt{2a\alpha q}e^{-i\varphi/2}$ (the so-called square-root topology). The real part of the energy is shown in Fig.~\ref{fig1}(b,c) as the $z$ coordinate. It determines the group velocity, which diverges as $1/\sqrt{q}$. Moreover, the group velocity is, in general, not aligned with the wave vector. We note that the imaginary part of the energy (shown in Fig.~\ref{fig1}(b,c) with false color) is complementary to the real part: $\Re E\sim\pm\sqrt{q}\cos\varphi/2$, $\Im E\sim\pm\sqrt{q}\sin\varphi/2$. It determines the decay or the growth of the corresponding states. The imaginary part is zero only along a single line in the parameter space (given by $\varphi=0$), ending at the exceptional point. For all other points, the states are either growing or decaying, which affects the behavior of WPs in dynamical experiments, as we will see below.

\emph{Circular-polarized WP.} Contrary to a diabolical point, associated with two degenerate eigenstates (which allows any linear combination of them to be a solution), an exceptional point corresponds to a single eigenstate. For the Hamiltonian \eqref{hamNH} at $q=0$, this eigenstate is $\ket{\psi_0}=(1,0)^T$: a circularly-polarized mode, typical for the Voigt points in optics. We  begin by considering a WP corresponding to the polarization of this eigenstate. It shows a  finite wave-vector distribution $\sigma_q$ (linked with the finite size of the exciting beam $\sigma_r=2\pi/\sigma_q$) centered on the exceptional point.
%Indeed, any experiment deals with finite size beams, represented by finite size WPs, and thus our description applies to any attempt to test the exceptional point.

\begin{figure}[tbp]
\includegraphics[width=1.0\linewidth]{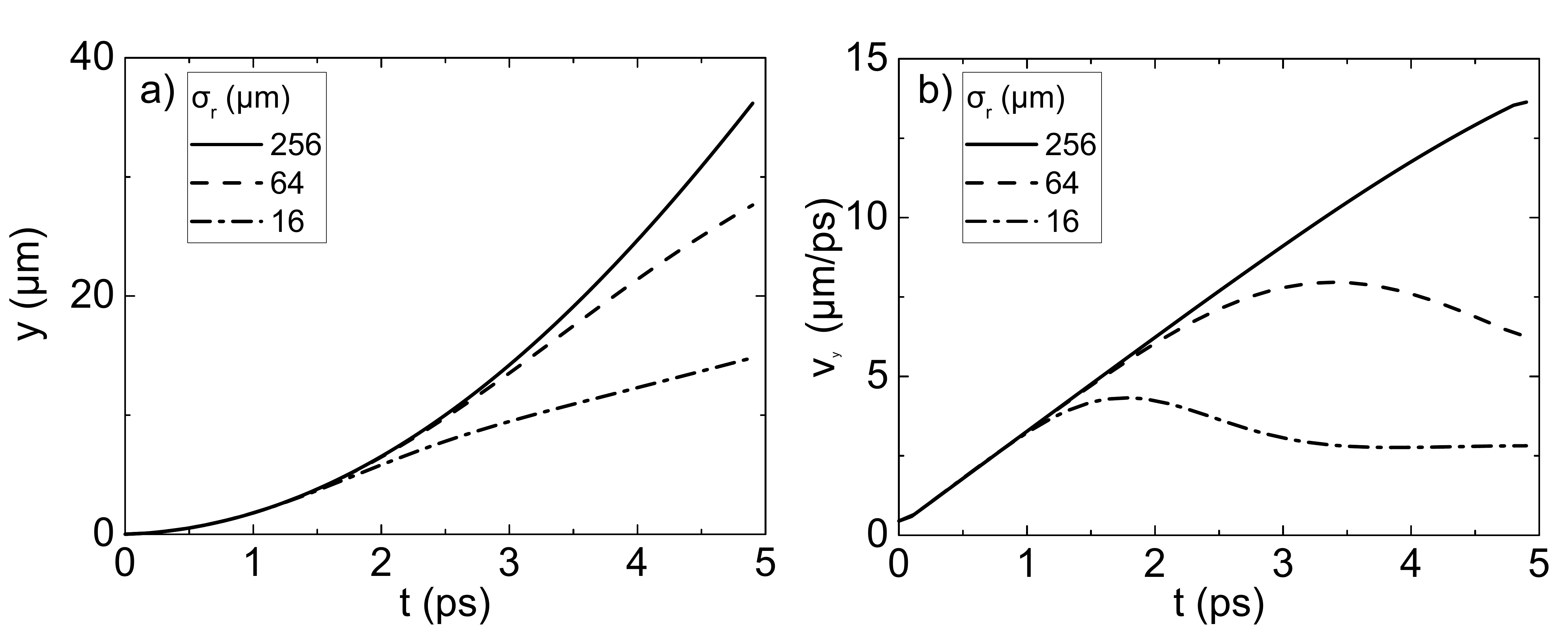}
 \caption{\label{fig2} Circular-polarized WP at an exceptional point. a) The center of mass position as a function of time. b) The center of mass velocity as a function of time, demonstrating constant acceleration. Solid, dashed, and dash-dotted lines correspond to different WP sizes.
 }
 \end{figure}
 
A WP dynamics depends on its projections on the eigenstates of the two branches. The most convenient tool to describe this projection in this multi-level system is the Eq.~\eqref{ovlint} with the metric tensor.
Let us start by calculating the radial component of the metric $g_{qq}$. For small $q$, the eigenstates of \eqref{hamNH} behave as $\ket{\psi_q}=(1-\alpha q/4a,e^{i\varphi/2}\sqrt{\alpha q/2a})^T$,
which allows to find:
\begin{equation}
    g_{qq}\approx\frac{\alpha^2}{16a^2}+\frac{\alpha}{8aq}
    \label{gqq}
\end{equation}
using Eq.~\eqref{qgt}. This $q^{-1}$ divergence of the metric is a general result, valid for any second-order exceptional point \cite{Brody2013}. We note that the divergence can be different for higher-order exceptional points, which will be discussed in a separate work. We also note that the other component of the metric tensor $g_{\varphi\varphi}$, whose divergence determines the dynamics of a WP center on a diabolical point, is now linear in $q$: $g_{\varphi\varphi}\approx\alpha q/8a$. Since the metric $g_{qq}$ does not depend on $\varphi$, the equation for the overlap integral between the states $\ket{\psi_0}$ and $\ket{\psi_q}$ can be explicitly written as:
\begin{equation}
    I_{circ}=1-\left(\int\limits_0^q \sqrt{g_{qq}~(dq)^2}\right)^2\neq f(\varphi)
\end{equation}
Since there is no dependence on $\varphi$, the WP is initially symmetrically distributed around the exceptional point on both branches. However, each individual state starts to grow or decline according to the imaginary part of its energy $\Im E$. The highest growth and decay rates are observed along the line $\varphi=\pi$, where $\sin\varphi/2=1$. The group velocity for all points along this line is perpendicular to the wave vector. It behaves as $v_y\sim q_x^{-1/2}$ (see \cite{suppl} for details). At the same time, the growth and decay rates of the bands behave as $\Gamma\sim\pm q_x^{1/2}$. The growing and decaying parts of the WP ($n_+$ and $n_-$) belong to different branches propagating in opposite directions along $y$. The average WP velocity for sufficiently small times and WP size is therefore given by
\begin{equation}
    \langle v_y \rangle = (n_+ -n_-) v_y\approx 2 v_y \Gamma t\approx 2\sqrt{2}\alpha a t
    \label{vycirc}
\end{equation}
which does not depend on the wave vector $q_x$, because the dependencies of $v_y$ and $\Gamma$ are inverse and compensate each other. We can therefore expect a finite-size WP centered at the eigenstate at the exceptional point to exhibit a constant acceleration in the vertical direction, because all of its components exhibit equal acceleration. This acceleration is proportional to the celerity $\alpha$ and the dichroism $a$.

This is confirmed by direct numerical simulations. We solve the time-dependent spinor Schr\"odinger equation with the Hamiltonian $\hat{H}_1$ defined by Eq.~\eqref{hamNH}
and extract the center of mass position as a function of time for different initial size of the WP, taking the parameters of the microcavities exhibiting a large birefringence (e.g. perovskite \cite{fieramosca2019chromodynamics}). 
The results of the simulations are shown in Fig.~\ref{fig2}. Panel (a) shows the evolution of the $y(t)$ coordinate of the center of mass as a function of time, which is clearly parabolic, and panel (b) shows the center of mass velocity $v_y(t)$, which grows linearly, as expected. The acceleration corresponds well to the analytical solution \eqref{vycirc}.  Changing the size of the WP in the real space $\sigma_r$ (as shown by the line style in Fig.~\ref{fig2}(a,b)) also leads to a very interesting and counter-intuitive behavior. Indeed, the linear increase of the velocity occurs only while the populations of the two branches $n_+$ and $n_-$ are comparable. The duration of this regime is determined by the maximal gain/loss ratio available within the WP size in the reciprocal space $\sigma_q$. For high $\sigma_q$, the regime of linear increase is lost more rapidly. While the wave vector of the center of mass $q_0$ of the WP is at this moment higher than for a smaller $\sigma_q$, the corresponding group velocity is lower, because $v_g\sim 1/\sqrt{q_0}$. So, a WP which is larger in reciprocal space (dash-dotted line) exhibits a smaller final velocity and a shorter acceleration period (and, finally, a smaller total displacement for the same amount of time).

\begin{figure}[tbp]
 \includegraphics[width=1.0\linewidth]{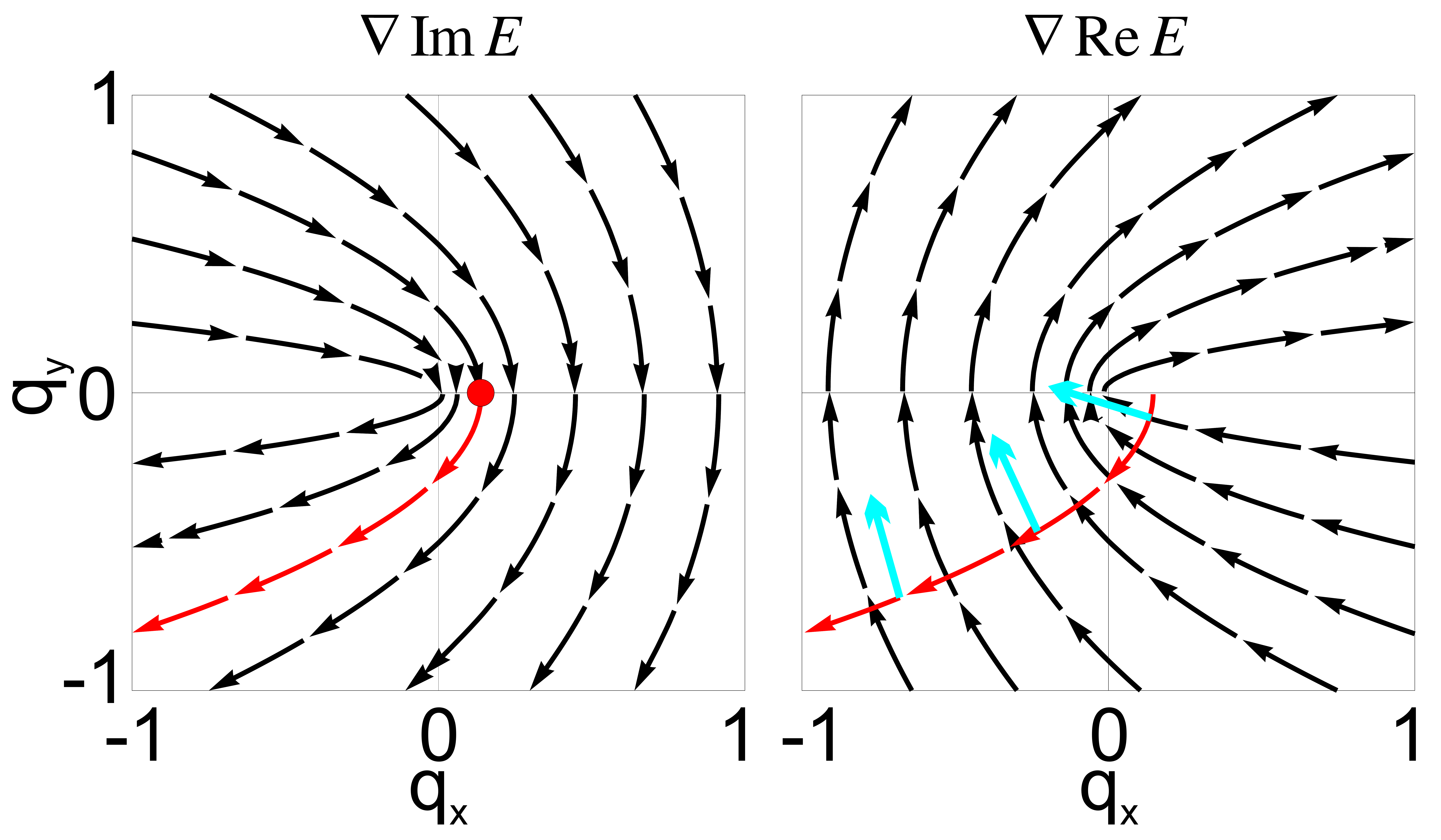}
  \caption{\label{fig3} a) The gradient of the imaginary part of the energy $\nabla\Im E$  for one dispersion branch. The arrows show how the center of mass of the WP moves in the reciprocal space due to the amplification of its components. The linear-polarized WP projection of the branch is marked with a red circle. b) The group velocity map $\nabla\Re E$ with the trajectory of the center of mass in red, and the corresponding group velocities highlighted in blue. 
 }
 \end{figure}

\emph{Linear-polarized WP.} As shown above, the quantum metric tensor $g_{qq}$ does not depend on $\varphi$, and thus the circular-polarized WP centered on the exceptional point exhibited exactly the same overlap with both branches at any $q$. Because of this, we had $n_+=n_-$ at $t=0$, and the initial WP velocity was zero.
The situation is different if the polarization of the WP does not correspond to the eigenstate of the exceptional point. As said above, the metric is divergent, which means that the eigenstates change \emph{very rapidly} with $q$. Because of this, even an infinitely small (in reciprocal space) linear-polarized WP exhibits a different overlap with the two branches. Moreover, its center of mass on each branch is not centered at $q=0$ any more. Indeed, using Eq.~\eqref{ovlint}, the overlap of the eigenstates with the H-polarized excitation spinor can be found as:
\begin{equation}
{I_{lin}} \approx \frac{1}{2} + { {\int\limits_{\left| {\psi \left( 0 \right)} \right\rangle }^{\left| {\psi \left( q \right)} \right\rangle } {\sqrt {{g_{qq}}} dq} }}
\approx \frac{1}{2}+\sqrt{\frac{\alpha q}{2a}}
\end{equation}
where we have used trigonometric identities to express the integral as a function of the metric at small $q>0$ ($I_{lin}=1/2$ for $q\leq 0$). The fast growth of the overlap leads to a non-zero effective center-of-mass wave vector $q_0$ for a Gaussian WP of any size, behaving as $q_0\sim\sigma_q^{3/2}\sqrt{\alpha/2a}$ for small $\sigma_q$ (large WPs in real space, small in reciprocal space). The position of this center of mass is determined by the polarization of the WP. This starting point is shown in Fig.~\ref{fig3} with a red circle.
The evolution of the WP over time in the reciprocal space is dictated by the imaginary part of the energy: the center of mass $\bm{q_0}$ shifts in the direction of the gradient $\nabla\Im E$, as shown in Fig.~\ref{fig3}(a) for a WP with initial H polarization (red arrows).  The ratio of the components of the gradient of the imaginary part of the energy $(\eta_x,\eta_y)=\nabla \Im E$ reads simply $\eta_y/\eta_x=(-q_x+\sqrt{q_x^2+q_y^2})/q_y$. For an initial wave vector $q_0$, this gives a parabolic trajectory $q_x=-(q_y^2-q_0^2)/(2q_0)$.  For sufficiently large $q_x$, one finds that $|q_y|\approx \sqrt{2q_0 |q_x|}$. 
The associated group velocity changes over time (blue arrows in Fig.~\ref{fig3}(b)), according to the position of the WP in the reciprocal space. The velocity of the very first moments in time corresponds to the high-energy terms of the Hamiltonian (celerity $\alpha$), which are determined by its $q$-dependent Hermitian part. Then, the velocity drops to the value determined by the ratio of the populations of the two bands $n_+$ and $n_-$, and the respective positions of the WP projections in the two bands. The $v_x$ projection of the group velocity for very small WPs (and $q_y\ll q_x$) is $v_x\approx3\sqrt{\alpha a}q_y^2/8q_x^2$ (see \cite{suppl}). To obtain the time dependence of the group velocity, we need to use the center of mass wave vector $q_x(t)$, whose calculation again involves the quantum metric via the overlap integral $I_{lin}$. For small $t$, $q_x(t)\approx \alpha a q_0^2 t^2/\sqrt{2\pi}$, and finally, combining all coefficients, the group velocity tends to a constant non-vanishing value \begin{equation}
\label{vx}
    \langle v_x \rangle \approx 0.38 \alpha
\end{equation}
which does not depend on the WP size. Moreover, it only depends on the celerity $\alpha$ of the Hermitian part of the Hamiltonian, and not on the non-Hermitian dichroism parameter $a$.

\begin{figure}[tbp]
 \includegraphics[width=1.0\linewidth]{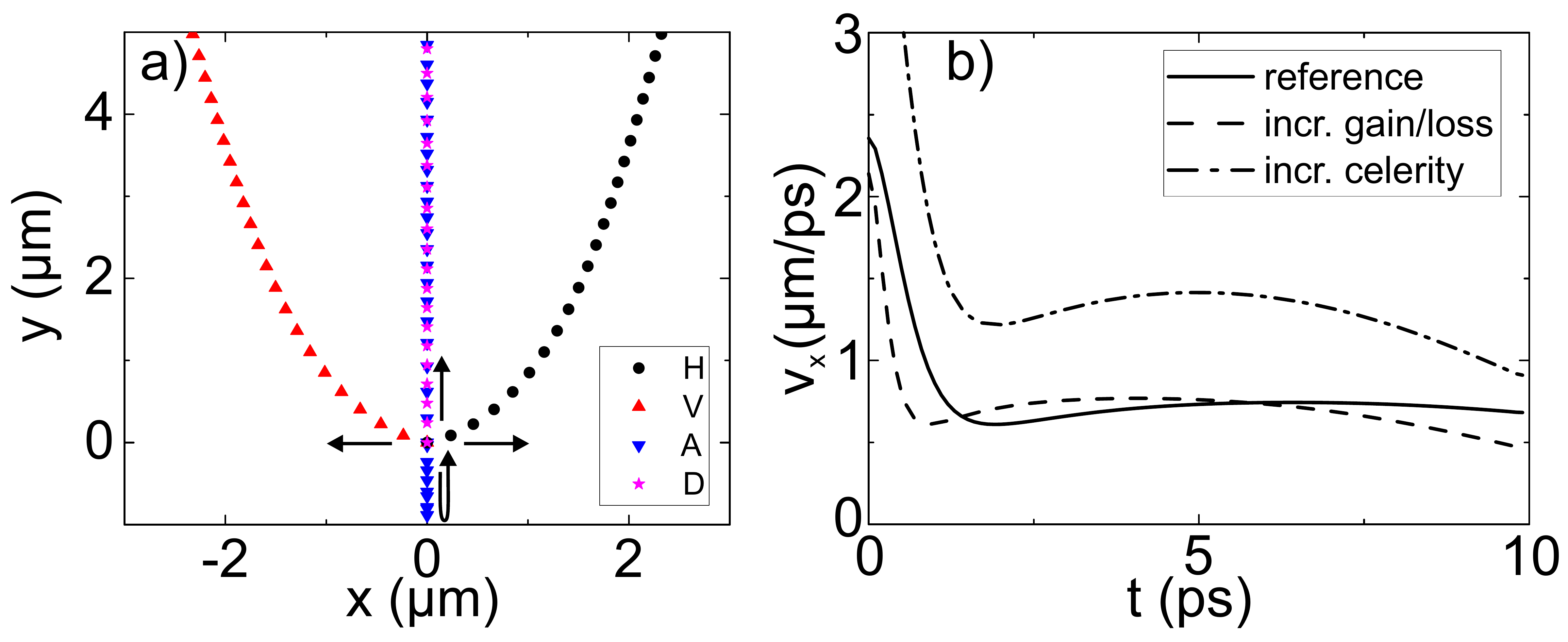}
  \caption{\label{fig4} a) Polarization-dependent trajectories for WPs centered at an exceptional point. b) The $x$-projection of the H-polarized WP velocity over time (solid curve -- same parameters as in (a); dashed curve -- increased $a$ (gain/losses); dash-dotted curve -- increased celerity $\alpha$).
 }
 \end{figure}

All this is indeed confirmed by numerical simulations. Figure~\ref{fig4}(a) shows all possible trajectories for WPs of different polarizations. They all demonstrate a constant acceleration along $y$, as for a circular WP. An additional polarization dependent constant velocity contribution appears. For an H-polarized WP (black circles in panel (a)), it is directed along $x$ and therefore can be studied independently from the other contribution $v_y$. Fig.~\ref{fig4}(b) shows that the velocity $v_x$ of an H-polarized WP quickly drops from the value given by the Hermitian part of the Hamiltonian ($v_x=\alpha$) at $t=0$ down to the constant value predicted by \eqref{vx} and indeed independent of the dichroism $a$ (dashed line). This finite constant velocity differs drastically from the behavoir of a gapped Dirac Hamiltonian, where the radial metric decays as $g_{qq}\sim -q^2$ and therefore does not diverge. Because of this, the associated group velocity tends to zero in the limit of a WP infinitely large in real space $v_{Dirac}\sim 1/\sigma_r^2\to 0$.  Contrary to a diabolical point, associated with a localized Berry curvature (delta function), an exceptional point exhibits distributed Berry curvature \cite{PhysRevA.86.064104,PhysRevA.99.032121}.  However, dynamical effects associated with this Berry curvature, such as the anomalous Hall effect, are practically unobservable, because the divergent group velocity dominates all possible corrections.

The effects we have described can be measured optically in microcavities with exceptional points \cite{Richter2019}, in atomic vapors \cite{Zhang2016}, or in dichroic birefringent crystals \cite{Berry2003}, but the theory we developed applies to all 2nd-order exceptional points. We note that contrary to the anomalous Hall effect, recently observed in microcavities \cite{gianfrate2020measurement}, the WP displacement does not saturate and continues to grow with time or propagation distance.

 These results demonstrate that the quantum metric plays a particularly important role in vicinity of the exceptional points, determining the dynamical behavior of WPs. The crucial feature is that the radial component of the quantum metric diverges. This, together with the divergent group velocity, leads to a non-vanishing polarization-dependent velocity for any finite-size WP, centered at the exceptional point. Our studies are important for future research and applications in non-Hermitian photonics.

\begin{acknowledgments}
We acknowledge the support of the projects EU "QUANTOPOL" (846353), "Quantum Fluids of Light"  (ANR-16-CE30-0021), of the ANR Labex GaNEXT (ANR-11-LABX-0014), and of the ANR program "Investissements d'Avenir" through the IDEX-ISITE initiative 16-IDEX-0001 (CAP 20-25). 
\end{acknowledgments}

\section{Supplemental Materials}

In this Supplemental material, we provide more details on the derivation of the expressions for the group velocity and the wave packet wave vector for circular and linear polarization. We also show the additional results of numerical simulations that confirm our theoretical analysis.

\subsection{Circular wavepacket}

To find this group velocity, one has first to determine the dispersion, then to take its derivative, and then to fix the wave vector, at which the value of the group velocity is needed. We therefore keep both components of the wave vector as variables at first, and only in the end of the calculation set $q_y=0$.
We write the series expansion of the real part of the energy for small $q$ along the line $\varphi=\pi$, where we use the identity $\cos\varphi/2=\sqrt{(1+\cos\varphi)/2}$:
\begin{equation}
    E(\bm{q})\approx \sqrt{2\alpha a} \left(q_x^2+q_y^2\right)^{1/4}\sqrt{\frac{1+\frac{q_x}{\sqrt{q_x^2+q_y^2}}}{2}}\approx \sqrt{\alpha a}\frac{q_y}{|q_x|^{1/2}}    
    \label{energy}
\end{equation}
Here, we have first used that $q$ is small, and then that $q_y\ll q_x$, and also that $|q_x|=-q_x$ for $q_x<0$. The group velocity in the vertical direction is given by $v_y=\partial E/\partial q_y$, which gives the expression
\begin{equation}
v_y\approx \frac{\sqrt{\alpha a}}{|q_x|^{1/2}}
\end{equation}
used in the main text.

\begin{figure*}[tbp]
\includegraphics[width=1.0\linewidth]{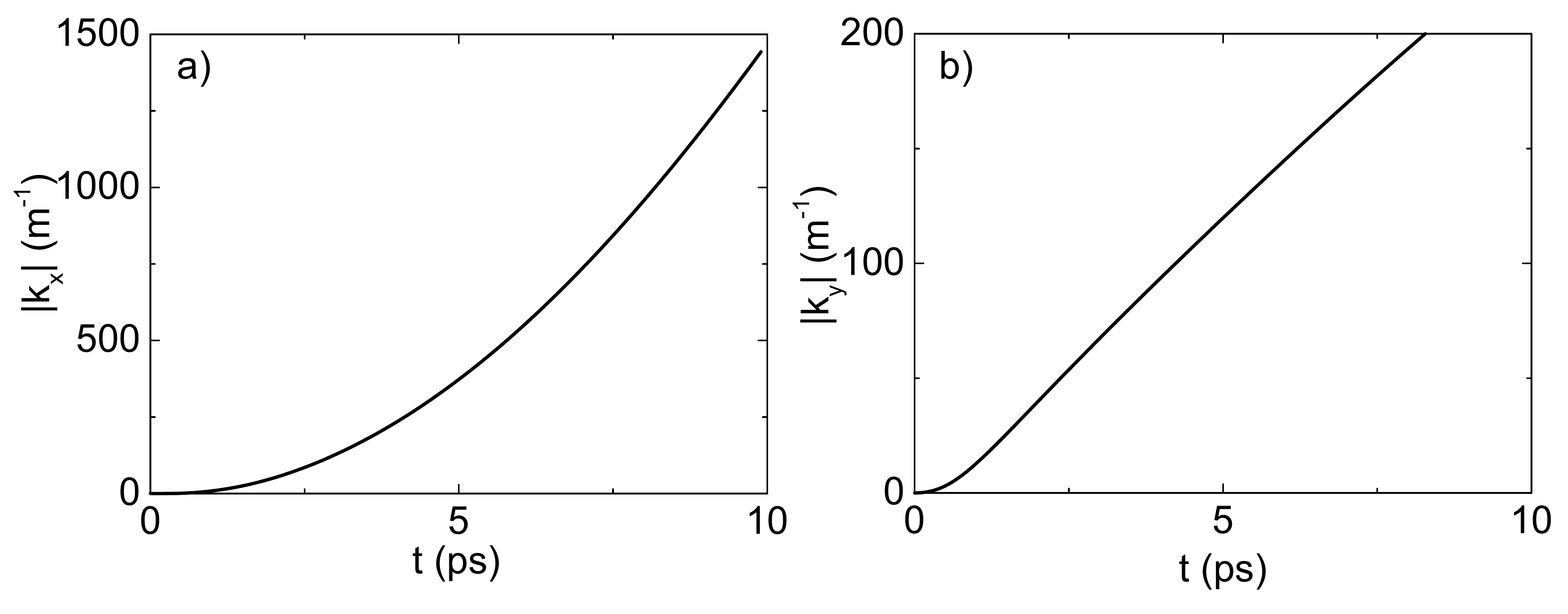}
 \caption{\label{figS1} The center of mass wave vector as a function of time, obtained from numerical simulations for a linearly-polarized wave packet: a) $k_x(t)$ (parabolic growth), b) $k_y(t)$ (linear growth).
 }
 \end{figure*}

\subsection{Linearly-polarized wavepacket}

The effective center of mass wave vector for one of the branches at $t=0$ is determined as:
\begin{equation}
q_0=\int q\times I_{lin}\times |\psi_0(q)|^2~dq
\end{equation}
where $|\psi_0(q)|^2$ is the excitation wavefunction (Gaussian WP with $\sigma_q$ width), $I_{lin}$ is the overlap integral with the chosen branch (given in the main text). The integral has to be divided into two parts, $q\leq 0$ and $q>0$. The contributions with $I_{lin}=1/2$ compensate each other, while the remaining term for $q>0$ gives the dependence $q_0\sim \sigma_q^{3/2}$ indicated in the main text.

To find the expression for the group velocity projection on the $x$ axis, we use the expression for the real part of the energy once again
\begin{equation}
    E(\bm{q})\approx \sqrt{2\alpha a} \left(q_x^2+q_y^2\right)^{1/4}\sqrt{\frac{1+\frac{q_x}{\sqrt{q_x^2+q_y^2}}}{2}}   
    \label{energy2}
\end{equation}
which we derive over $q_y$ and then apply series expansion over $q_y$, taking care of $q_x<0$. This gives the expression for $v_x\approx3\sqrt{\alpha a}q_y^2/8q_x^2$ given in the main text.

For the calculation of $q_x(t)$ (the $x$ projection of the effective center of mass coordinate), we take into account that the difference of the population of the branches is controlled by the overlap integral, which behaves as $I_{lin}\approx 1/2+\sqrt{\alpha q_x/2a}$ and by the imaginary part $\Gamma t\sim\sqrt{q_x} t$. The position of the center of mass on a given branch is therefore given by
\begin{equation}
\langle q_1 \rangle \sim \int q (1+q t) |\psi_0(q)|^2~dq\sim \sigma_q^2 t
\end{equation}
while the population of this branch is given by
\begin{equation}
I_1\sim \int (1+q t)|\psi_0(q)|^2~dq\sim \sigma_q t
\end{equation}
while the sum of the two branches $I_1+I_2$ remains approximately constant (one branch grows, the other decays). We therefore find that 
\begin{equation}
q_x(t)\sim \sigma_q^3 t^2
\end{equation}
where $\sigma_q^3 \sim q_0^2$ (the initial wave vector of a WP within one of the branches).
This allows to find $v_x$ using $q_x(t)$ and $q_y=\sqrt{2q_0q_x}$.

The results concerning the dynamics of a linear-polarized wavepacket, provided in the main text, are strongly based on its center of mass position in the reciprocal space. As an additional confirmation of our theoretical considerations, we provide the numerical results on the center of mass wave vector in this supplemental material. Figure~\ref{figS1} shows the two projections of the center of mass wavevector as a function of time. Panel (a) clearly exhibits a parabolic growth (which is valid both for circular and linear wavepackets), whereas panel (b) demonstrates a linear dependence on time. This is due to the fact that the center of mass wave vector follows the trajectories of maximal gradient of the imaginary part of the energy shown in Fig.~3 of the main text, which represent a family of parabolic curves, and therefore $k_y\sim\sqrt{k_x}\sim t$. The parameters were the same as in Fig.~4 of the main text.

%\bibliography{biblio}

\begin{thebibliography}{44}
\expandafter\ifx\csname natexlab\endcsname\relax\def\natexlab#1{#1}\fi
\expandafter\ifx\csname bibnamefont\endcsname\relax
  \def\bibnamefont#1{#1}\fi
\expandafter\ifx\csname bibfnamefont\endcsname\relax
  \def\bibfnamefont#1{#1}\fi
\expandafter\ifx\csname citenamefont\endcsname\relax
  \def\citenamefont#1{#1}\fi
\expandafter\ifx\csname url\endcsname\relax
  \def\url#1{\texttt{#1}}\fi
\expandafter\ifx\csname urlprefix\endcsname\relax\def\urlprefix{URL }\fi
\providecommand{\bibinfo}[2]{#2}
\providecommand{\eprint}[2][]{\url{#2}}

\bibitem[{\citenamefont{Moiseyev}(2011)}]{Moiseyev2011}
\bibinfo{author}{\bibfnamefont{N.}~\bibnamefont{Moiseyev}},
  \emph{\bibinfo{title}{Non-Hermitian quantum mechanics}}
  (\bibinfo{publisher}{Cambridge University Press (Cambridge, UK)},
  \bibinfo{year}{2011}).

\bibitem[{\citenamefont{{\"O}zdemir et~al.}(2019)\citenamefont{{\"O}zdemir,
  Rotter, Nori, and Yang}}]{Ozdemir2019}
\bibinfo{author}{\bibfnamefont{{\c{S}}.}~\bibnamefont{{\"O}zdemir}},
  \bibinfo{author}{\bibfnamefont{S.}~\bibnamefont{Rotter}},
  \bibinfo{author}{\bibfnamefont{F.}~\bibnamefont{Nori}}, \bibnamefont{and}
  \bibinfo{author}{\bibfnamefont{L.}~\bibnamefont{Yang}},
  \bibinfo{journal}{Nature materials} \textbf{\bibinfo{volume}{18}},
  \bibinfo{pages}{783} (\bibinfo{year}{2019}).

\bibitem[{\citenamefont{Konotop et~al.}(2016)\citenamefont{Konotop, Yang, and
  Zezyulin}}]{Konotop2016}
\bibinfo{author}{\bibfnamefont{V.~V.} \bibnamefont{Konotop}},
  \bibinfo{author}{\bibfnamefont{J.}~\bibnamefont{Yang}}, \bibnamefont{and}
  \bibinfo{author}{\bibfnamefont{D.~A.} \bibnamefont{Zezyulin}},
  \bibinfo{journal}{Rev. Mod. Phys.} \textbf{\bibinfo{volume}{88}},
  \bibinfo{pages}{035002} (\bibinfo{year}{2016}),
  \urlprefix\url{https://link.aps.org/doi/10.1103/RevModPhys.88.035002}.

\bibitem[{\citenamefont{El-Ganainy et~al.}(2018)\citenamefont{El-Ganainy,
  Makris, Khajavikhan, Musslimani, Rotter, and Christodoulides}}]{Ganainy2018}
\bibinfo{author}{\bibfnamefont{R.}~\bibnamefont{El-Ganainy}},
  \bibinfo{author}{\bibfnamefont{K.~G.} \bibnamefont{Makris}},
  \bibinfo{author}{\bibfnamefont{M.}~\bibnamefont{Khajavikhan}},
  \bibinfo{author}{\bibfnamefont{Z.~H.} \bibnamefont{Musslimani}},
  \bibinfo{author}{\bibfnamefont{S.}~\bibnamefont{Rotter}}, \bibnamefont{and}
  \bibinfo{author}{\bibfnamefont{D.~N.} \bibnamefont{Christodoulides}},
  \bibinfo{journal}{Nature Physics} \textbf{\bibinfo{volume}{14}},
  \bibinfo{pages}{11} (\bibinfo{year}{2018}).

\bibitem[{\citenamefont{Leykam et~al.}(2017)\citenamefont{Leykam, Bliokh,
  Huang, Chong, and Nori}}]{Bliokh2017}
\bibinfo{author}{\bibfnamefont{D.}~\bibnamefont{Leykam}},
  \bibinfo{author}{\bibfnamefont{K.~Y.} \bibnamefont{Bliokh}},
  \bibinfo{author}{\bibfnamefont{C.}~\bibnamefont{Huang}},
  \bibinfo{author}{\bibfnamefont{Y.~D.} \bibnamefont{Chong}}, \bibnamefont{and}
  \bibinfo{author}{\bibfnamefont{F.}~\bibnamefont{Nori}},
  \bibinfo{journal}{Phys. Rev. Lett.} \textbf{\bibinfo{volume}{118}},
  \bibinfo{pages}{040401} (\bibinfo{year}{2017}),
  \urlprefix\url{https://link.aps.org/doi/10.1103/PhysRevLett.118.040401}.

\bibitem[{\citenamefont{Cui and Zheng}(2012)}]{PhysRevA.86.064104}
\bibinfo{author}{\bibfnamefont{X.-D.} \bibnamefont{Cui}} \bibnamefont{and}
  \bibinfo{author}{\bibfnamefont{Y.}~\bibnamefont{Zheng}},
  \bibinfo{journal}{Phys. Rev. A} \textbf{\bibinfo{volume}{86}},
  \bibinfo{pages}{064104} (\bibinfo{year}{2012}),
  \urlprefix\url{https://link.aps.org/doi/10.1103/PhysRevA.86.064104}.

\bibitem[{\citenamefont{Zhang and Wu}(2019)}]{PhysRevA.99.032121}
\bibinfo{author}{\bibfnamefont{Q.}~\bibnamefont{Zhang}} \bibnamefont{and}
  \bibinfo{author}{\bibfnamefont{B.}~\bibnamefont{Wu}}, \bibinfo{journal}{Phys.
  Rev. A} \textbf{\bibinfo{volume}{99}}, \bibinfo{pages}{032121}
  (\bibinfo{year}{2019}),
  \urlprefix\url{https://link.aps.org/doi/10.1103/PhysRevA.99.032121}.

\bibitem[{\citenamefont{Brody and Graefe}(2013)}]{Brody2013}
\bibinfo{author}{\bibfnamefont{D.~C.} \bibnamefont{Brody}} \bibnamefont{and}
  \bibinfo{author}{\bibfnamefont{E.~M.} \bibnamefont{Graefe}},
  \bibinfo{journal}{Entropy} \textbf{\bibinfo{volume}{15}},
  \bibinfo{pages}{3361} (\bibinfo{year}{2013}), ISSN \bibinfo{issn}{10994300},
  \eprint{1307.4017}.

\bibitem[{\citenamefont{Provost and Vallee}(1980)}]{provost1980riemannian}
\bibinfo{author}{\bibfnamefont{J.}~\bibnamefont{Provost}} \bibnamefont{and}
  \bibinfo{author}{\bibfnamefont{G.}~\bibnamefont{Vallee}},
  \bibinfo{journal}{Communications in Mathematical Physics}
  \textbf{\bibinfo{volume}{76}}, \bibinfo{pages}{289} (\bibinfo{year}{1980}).

\bibitem[{\citenamefont{Gianfrate et~al.}(2020)\citenamefont{Gianfrate, Bleu,
  Dominici, Ardizzone, De~Giorgi, Ballarini, Lerario, West, Pfeiffer,
  Solnyshkov et~al.}}]{gianfrate2020measurement}
\bibinfo{author}{\bibfnamefont{A.}~\bibnamefont{Gianfrate}},
  \bibinfo{author}{\bibfnamefont{O.}~\bibnamefont{Bleu}},
  \bibinfo{author}{\bibfnamefont{L.}~\bibnamefont{Dominici}},
  \bibinfo{author}{\bibfnamefont{V.}~\bibnamefont{Ardizzone}},
  \bibinfo{author}{\bibfnamefont{M.}~\bibnamefont{De~Giorgi}},
  \bibinfo{author}{\bibfnamefont{D.}~\bibnamefont{Ballarini}},
  \bibinfo{author}{\bibfnamefont{G.}~\bibnamefont{Lerario}},
  \bibinfo{author}{\bibfnamefont{K.}~\bibnamefont{West}},
  \bibinfo{author}{\bibfnamefont{L.}~\bibnamefont{Pfeiffer}},
  \bibinfo{author}{\bibfnamefont{D.}~\bibnamefont{Solnyshkov}},
  \bibnamefont{et~al.}, \bibinfo{journal}{Nature}
  \textbf{\bibinfo{volume}{578}}, \bibinfo{pages}{381} (\bibinfo{year}{2020}).

\bibitem[{\citenamefont{Yu et~al.}(2019)\citenamefont{Yu, Yang, Gong, Cao, Lu,
  Liu, Zhang, Plenio, Jelezko, Ozawa et~al.}}]{Yu2019}
\bibinfo{author}{\bibfnamefont{M.}~\bibnamefont{Yu}},
  \bibinfo{author}{\bibfnamefont{P.}~\bibnamefont{Yang}},
  \bibinfo{author}{\bibfnamefont{M.}~\bibnamefont{Gong}},
  \bibinfo{author}{\bibfnamefont{Q.}~\bibnamefont{Cao}},
  \bibinfo{author}{\bibfnamefont{Q.}~\bibnamefont{Lu}},
  \bibinfo{author}{\bibfnamefont{H.}~\bibnamefont{Liu}},
  \bibinfo{author}{\bibfnamefont{S.}~\bibnamefont{Zhang}},
  \bibinfo{author}{\bibfnamefont{M.~B.} \bibnamefont{Plenio}},
  \bibinfo{author}{\bibfnamefont{F.}~\bibnamefont{Jelezko}},
  \bibinfo{author}{\bibfnamefont{T.}~\bibnamefont{Ozawa}},
  \bibnamefont{et~al.}, \bibinfo{journal}{National Science Review}
  \textbf{\bibinfo{volume}{7}}, \bibinfo{pages}{254} (\bibinfo{year}{2019}),
  ISSN \bibinfo{issn}{2095-5138},
  \eprint{https://academic.oup.com/nsr/article-pdf/7/2/254/32921980/nwz193.pdf},
  \urlprefix\url{https://doi.org/10.1093/nsr/nwz193}.

\bibitem[{\citenamefont{Zanardi et~al.}(2007)\citenamefont{Zanardi, Giorda, and
  Cozzini}}]{Zanardi2007}
\bibinfo{author}{\bibfnamefont{P.}~\bibnamefont{Zanardi}},
  \bibinfo{author}{\bibfnamefont{P.}~\bibnamefont{Giorda}}, \bibnamefont{and}
  \bibinfo{author}{\bibfnamefont{M.}~\bibnamefont{Cozzini}},
  \bibinfo{journal}{Phys. Rev. Lett.} \textbf{\bibinfo{volume}{99}},
  \bibinfo{pages}{100603} (\bibinfo{year}{2007}),
  \urlprefix\url{https://link.aps.org/doi/10.1103/PhysRevLett.99.100603}.

\bibitem[{\citenamefont{Gao et~al.}(2014)\citenamefont{Gao, Yang, and
  Niu}}]{Gao2014}
\bibinfo{author}{\bibfnamefont{Y.}~\bibnamefont{Gao}},
  \bibinfo{author}{\bibfnamefont{S.~A.} \bibnamefont{Yang}}, \bibnamefont{and}
  \bibinfo{author}{\bibfnamefont{Q.}~\bibnamefont{Niu}},
  \bibinfo{journal}{Phys. Rev. Lett.} \textbf{\bibinfo{volume}{112}},
  \bibinfo{pages}{166601} (\bibinfo{year}{2014}),
  \urlprefix\url{https://link.aps.org/doi/10.1103/PhysRevLett.112.166601}.

\bibitem[{\citenamefont{Pi\'echon et~al.}(2016)\citenamefont{Pi\'echon, Raoux,
  Fuchs, and Montambaux}}]{Piechon2016}
\bibinfo{author}{\bibfnamefont{F.}~\bibnamefont{Pi\'echon}},
  \bibinfo{author}{\bibfnamefont{A.}~\bibnamefont{Raoux}},
  \bibinfo{author}{\bibfnamefont{J.-N.} \bibnamefont{Fuchs}}, \bibnamefont{and}
  \bibinfo{author}{\bibfnamefont{G.}~\bibnamefont{Montambaux}},
  \bibinfo{journal}{Phys. Rev. B} \textbf{\bibinfo{volume}{94}},
  \bibinfo{pages}{134423} (\bibinfo{year}{2016}),
  \urlprefix\url{https://link.aps.org/doi/10.1103/PhysRevB.94.134423}.

\bibitem[{\citenamefont{Srivastava and
  Imamoglu}(2015)}]{PhysRevLett.115.166802}
\bibinfo{author}{\bibfnamefont{A.}~\bibnamefont{Srivastava}} \bibnamefont{and}
  \bibinfo{author}{\bibfnamefont{A.}~\bibnamefont{Imamoglu}},
  \bibinfo{journal}{Phys. Rev. Lett.} \textbf{\bibinfo{volume}{115}},
  \bibinfo{pages}{166802} (\bibinfo{year}{2015}),
  \urlprefix\url{http://link.aps.org/doi/10.1103/PhysRevLett.115.166802}.

\bibitem[{\citenamefont{Peotta and
  T{\"o}rm{\"a}}(2015)}]{peotta2015superfluidity}
\bibinfo{author}{\bibfnamefont{S.}~\bibnamefont{Peotta}} \bibnamefont{and}
  \bibinfo{author}{\bibfnamefont{P.}~\bibnamefont{T{\"o}rm{\"a}}},
  \bibinfo{journal}{Nature communications} \textbf{\bibinfo{volume}{6}},
  \bibinfo{pages}{8944} (\bibinfo{year}{2015}).

\bibitem[{\citenamefont{Liang et~al.}(2017)\citenamefont{Liang, Peotta, Harju,
  and T\"orm\"a}}]{Liang2017}
\bibinfo{author}{\bibfnamefont{L.}~\bibnamefont{Liang}},
  \bibinfo{author}{\bibfnamefont{S.}~\bibnamefont{Peotta}},
  \bibinfo{author}{\bibfnamefont{A.}~\bibnamefont{Harju}}, \bibnamefont{and}
  \bibinfo{author}{\bibfnamefont{P.}~\bibnamefont{T\"orm\"a}},
  \bibinfo{journal}{Phys. Rev. B} \textbf{\bibinfo{volume}{96}},
  \bibinfo{pages}{064511} (\bibinfo{year}{2017}),
  \urlprefix\url{https://link.aps.org/doi/10.1103/PhysRevB.96.064511}.

\bibitem[{\citenamefont{Bleu et~al.}(2018)\citenamefont{Bleu, Malpuech, Gao,
  and Solnyshkov}}]{Bleu2018effective}
\bibinfo{author}{\bibfnamefont{O.}~\bibnamefont{Bleu}},
  \bibinfo{author}{\bibfnamefont{G.}~\bibnamefont{Malpuech}},
  \bibinfo{author}{\bibfnamefont{Y.}~\bibnamefont{Gao}}, \bibnamefont{and}
  \bibinfo{author}{\bibfnamefont{D.~D.} \bibnamefont{Solnyshkov}},
  \bibinfo{journal}{Phys. Rev. Lett.} \textbf{\bibinfo{volume}{121}},
  \bibinfo{pages}{020401} (\bibinfo{year}{2018}),
  \urlprefix\url{https://link.aps.org/doi/10.1103/PhysRevLett.121.020401}.

\bibitem[{\citenamefont{Shen et~al.}(2018)\citenamefont{Shen, Zhen, and
  Fu}}]{Shen2018}
\bibinfo{author}{\bibfnamefont{H.}~\bibnamefont{Shen}},
  \bibinfo{author}{\bibfnamefont{B.}~\bibnamefont{Zhen}}, \bibnamefont{and}
  \bibinfo{author}{\bibfnamefont{L.}~\bibnamefont{Fu}}, \bibinfo{journal}{Phys.
  Rev. Lett.} \textbf{\bibinfo{volume}{120}}, \bibinfo{pages}{146402}
  (\bibinfo{year}{2018}),
  \urlprefix\url{https://link.aps.org/doi/10.1103/PhysRevLett.120.146402}.

\bibitem[{\citenamefont{Zhong et~al.}(2018)\citenamefont{Zhong, Khajavikhan,
  Christodoulides, and El-Ganainy}}]{zhong2018winding}
\bibinfo{author}{\bibfnamefont{Q.}~\bibnamefont{Zhong}},
  \bibinfo{author}{\bibfnamefont{M.}~\bibnamefont{Khajavikhan}},
  \bibinfo{author}{\bibfnamefont{D.~N.} \bibnamefont{Christodoulides}},
  \bibnamefont{and}
  \bibinfo{author}{\bibfnamefont{R.}~\bibnamefont{El-Ganainy}},
  \bibinfo{journal}{Nature communications} \textbf{\bibinfo{volume}{9}},
  \bibinfo{pages}{1} (\bibinfo{year}{2018}).

\bibitem[{\citenamefont{Berry and Uzdin}(2011)}]{Berry2011a}
\bibinfo{author}{\bibfnamefont{M.}~\bibnamefont{Berry}} \bibnamefont{and}
  \bibinfo{author}{\bibfnamefont{R.}~\bibnamefont{Uzdin}},
  \bibinfo{journal}{Journal of Physics A: Mathematical and Theoretical}
  \textbf{\bibinfo{volume}{44}}, \bibinfo{pages}{435303}
  (\bibinfo{year}{2011}).

\bibitem[{\citenamefont{Berry}(2011)}]{Berry2011b}
\bibinfo{author}{\bibfnamefont{M.}~\bibnamefont{Berry}},
  \bibinfo{journal}{Journal of Optics} \textbf{\bibinfo{volume}{13}},
  \bibinfo{pages}{115701} (\bibinfo{year}{2011}).

\bibitem[{\citenamefont{Milburn et~al.}(2015)\citenamefont{Milburn, Doppler,
  Holmes, Portolan, Rotter, and Rabl}}]{Milburn2015}
\bibinfo{author}{\bibfnamefont{T.~J.} \bibnamefont{Milburn}},
  \bibinfo{author}{\bibfnamefont{J.}~\bibnamefont{Doppler}},
  \bibinfo{author}{\bibfnamefont{C.~A.} \bibnamefont{Holmes}},
  \bibinfo{author}{\bibfnamefont{S.}~\bibnamefont{Portolan}},
  \bibinfo{author}{\bibfnamefont{S.}~\bibnamefont{Rotter}}, \bibnamefont{and}
  \bibinfo{author}{\bibfnamefont{P.}~\bibnamefont{Rabl}},
  \bibinfo{journal}{Phys. Rev. A} \textbf{\bibinfo{volume}{92}},
  \bibinfo{pages}{052124} (\bibinfo{year}{2015}),
  \urlprefix\url{https://link.aps.org/doi/10.1103/PhysRevA.92.052124}.

\bibitem[{\citenamefont{Doppler et~al.}(2016)\citenamefont{Doppler, Mailybaev,
  B{\"o}hm, Kuhl, Girschik, Libisch, Milburn, Rabl, Moiseyev, and
  Rotter}}]{Doppler2016}
\bibinfo{author}{\bibfnamefont{J.}~\bibnamefont{Doppler}},
  \bibinfo{author}{\bibfnamefont{A.~A.} \bibnamefont{Mailybaev}},
  \bibinfo{author}{\bibfnamefont{J.}~\bibnamefont{B{\"o}hm}},
  \bibinfo{author}{\bibfnamefont{U.}~\bibnamefont{Kuhl}},
  \bibinfo{author}{\bibfnamefont{A.}~\bibnamefont{Girschik}},
  \bibinfo{author}{\bibfnamefont{F.}~\bibnamefont{Libisch}},
  \bibinfo{author}{\bibfnamefont{T.~J.} \bibnamefont{Milburn}},
  \bibinfo{author}{\bibfnamefont{P.}~\bibnamefont{Rabl}},
  \bibinfo{author}{\bibfnamefont{N.}~\bibnamefont{Moiseyev}}, \bibnamefont{and}
  \bibinfo{author}{\bibfnamefont{S.}~\bibnamefont{Rotter}},
  \bibinfo{journal}{Nature} \textbf{\bibinfo{volume}{537}}, \bibinfo{pages}{76}
  (\bibinfo{year}{2016}).

\bibitem[{\citenamefont{Zhang et~al.}(2018)\citenamefont{Zhang, Wang, Hou, and
  Chan}}]{Zhang2018}
\bibinfo{author}{\bibfnamefont{X.~L.} \bibnamefont{Zhang}},
  \bibinfo{author}{\bibfnamefont{S.}~\bibnamefont{Wang}},
  \bibinfo{author}{\bibfnamefont{B.}~\bibnamefont{Hou}}, \bibnamefont{and}
  \bibinfo{author}{\bibfnamefont{C.~T.} \bibnamefont{Chan}},
  \bibinfo{journal}{Physical Review X} \textbf{\bibinfo{volume}{8}},
  \bibinfo{pages}{21066} (\bibinfo{year}{2018}), ISSN \bibinfo{issn}{21603308},
  \eprint{1804.09145},
  \urlprefix\url{https://doi.org/10.1103/PhysRevX.8.021066}.

\bibitem[{\citenamefont{Berry and Shukla}(2020)}]{Berry2020}
\bibinfo{author}{\bibfnamefont{M.~V.} \bibnamefont{Berry}} \bibnamefont{and}
  \bibinfo{author}{\bibfnamefont{P.}~\bibnamefont{Shukla}},
  \bibinfo{journal}{Journal of Physics A: Mathematical and Theoretical}
  \textbf{\bibinfo{volume}{53}}, \bibinfo{pages}{275202}
  (\bibinfo{year}{2020}),
  \urlprefix\url{https://doi.org/10.1088%2F1751-8121%2Fab91d6}.

\bibitem[{\citenamefont{Hahn et~al.}(2016)\citenamefont{Hahn, Choi, Yoon, Song,
  Oh, and Berini}}]{Hahn2016}
\bibinfo{author}{\bibfnamefont{C.}~\bibnamefont{Hahn}},
  \bibinfo{author}{\bibfnamefont{Y.}~\bibnamefont{Choi}},
  \bibinfo{author}{\bibfnamefont{J.~W.} \bibnamefont{Yoon}},
  \bibinfo{author}{\bibfnamefont{S.~H.} \bibnamefont{Song}},
  \bibinfo{author}{\bibfnamefont{C.~H.} \bibnamefont{Oh}}, \bibnamefont{and}
  \bibinfo{author}{\bibfnamefont{P.}~\bibnamefont{Berini}},
  \bibinfo{journal}{Nature Communications} \textbf{\bibinfo{volume}{7}},
  \bibinfo{pages}{1} (\bibinfo{year}{2016}), ISSN \bibinfo{issn}{20411723}.

\bibitem[{\citenamefont{Zhang et~al.}(2016)\citenamefont{Zhang, Zhang, Sheng,
  Yang, Miri, Christodoulides, He, Zhang, and Xiao}}]{Zhang2016}
\bibinfo{author}{\bibfnamefont{Z.}~\bibnamefont{Zhang}},
  \bibinfo{author}{\bibfnamefont{Y.}~\bibnamefont{Zhang}},
  \bibinfo{author}{\bibfnamefont{J.}~\bibnamefont{Sheng}},
  \bibinfo{author}{\bibfnamefont{L.}~\bibnamefont{Yang}},
  \bibinfo{author}{\bibfnamefont{M.-A.} \bibnamefont{Miri}},
  \bibinfo{author}{\bibfnamefont{D.~N.} \bibnamefont{Christodoulides}},
  \bibinfo{author}{\bibfnamefont{B.}~\bibnamefont{He}},
  \bibinfo{author}{\bibfnamefont{Y.}~\bibnamefont{Zhang}}, \bibnamefont{and}
  \bibinfo{author}{\bibfnamefont{M.}~\bibnamefont{Xiao}},
  \bibinfo{journal}{Phys. Rev. Lett.} \textbf{\bibinfo{volume}{117}},
  \bibinfo{pages}{123601} (\bibinfo{year}{2016}),
  \urlprefix\url{https://link.aps.org/doi/10.1103/PhysRevLett.117.123601}.

\bibitem[{\citenamefont{Wiersig}(2016)}]{Wiersig2016}
\bibinfo{author}{\bibfnamefont{J.}~\bibnamefont{Wiersig}},
  \bibinfo{journal}{Physical Review A} \textbf{\bibinfo{volume}{93}},
  \bibinfo{pages}{1} (\bibinfo{year}{2016}), ISSN \bibinfo{issn}{24699934}.

\bibitem[{\citenamefont{Naghiloo et~al.}(2019)\citenamefont{Naghiloo, Abbasi,
  Joglekar, and Murch}}]{Naghiloo2019}
\bibinfo{author}{\bibfnamefont{M.}~\bibnamefont{Naghiloo}},
  \bibinfo{author}{\bibfnamefont{M.}~\bibnamefont{Abbasi}},
  \bibinfo{author}{\bibfnamefont{Y.~N.} \bibnamefont{Joglekar}},
  \bibnamefont{and} \bibinfo{author}{\bibfnamefont{K.~W.} \bibnamefont{Murch}},
  \bibinfo{journal}{Nature Physics} \textbf{\bibinfo{volume}{15}},
  \bibinfo{pages}{1232} (\bibinfo{year}{2019}), ISSN \bibinfo{issn}{17452481},
  \eprint{1901.07968},
  \urlprefix\url{http://dx.doi.org/10.1038/s41567-019-0652-z}.

\bibitem[{\citenamefont{Kavokin et~al.}(2011)\citenamefont{Kavokin, Baumberg,
  Malpuech, and Laussy}}]{Microcavities}
\bibinfo{author}{\bibfnamefont{A.}~\bibnamefont{Kavokin}},
  \bibinfo{author}{\bibfnamefont{J.~J.} \bibnamefont{Baumberg}},
  \bibinfo{author}{\bibfnamefont{G.}~\bibnamefont{Malpuech}}, \bibnamefont{and}
  \bibinfo{author}{\bibfnamefont{F.~P.} \bibnamefont{Laussy}},
  \emph{\bibinfo{title}{Microcavities}} (\bibinfo{publisher}{Oxford University
  Press}, \bibinfo{year}{2011}).

\bibitem[{\citenamefont{Gao et~al.}(2015)\citenamefont{Gao, Estrecho, Bliokh,
  Liew, Fraser, Brodbeck, Kamp, Schneider, H{\"o}fling, Yamamoto
  et~al.}}]{Gao2015b}
\bibinfo{author}{\bibfnamefont{T.}~\bibnamefont{Gao}},
  \bibinfo{author}{\bibfnamefont{E.}~\bibnamefont{Estrecho}},
  \bibinfo{author}{\bibfnamefont{K.}~\bibnamefont{Bliokh}},
  \bibinfo{author}{\bibfnamefont{T.}~\bibnamefont{Liew}},
  \bibinfo{author}{\bibfnamefont{M.}~\bibnamefont{Fraser}},
  \bibinfo{author}{\bibfnamefont{S.}~\bibnamefont{Brodbeck}},
  \bibinfo{author}{\bibfnamefont{M.}~\bibnamefont{Kamp}},
  \bibinfo{author}{\bibfnamefont{C.}~\bibnamefont{Schneider}},
  \bibinfo{author}{\bibfnamefont{S.}~\bibnamefont{H{\"o}fling}},
  \bibinfo{author}{\bibfnamefont{Y.}~\bibnamefont{Yamamoto}},
  \bibnamefont{et~al.}, \bibinfo{journal}{Nature}
  \textbf{\bibinfo{volume}{526}}, \bibinfo{pages}{554} (\bibinfo{year}{2015}).

\bibitem[{\citenamefont{Gao et~al.}(2018)\citenamefont{Gao, Li, Estrecho, Liew,
  Comber-Todd, Nalitov, Steger, West, Pfeiffer, Snoke et~al.}}]{Gao2018}
\bibinfo{author}{\bibfnamefont{T.}~\bibnamefont{Gao}},
  \bibinfo{author}{\bibfnamefont{G.}~\bibnamefont{Li}},
  \bibinfo{author}{\bibfnamefont{E.}~\bibnamefont{Estrecho}},
  \bibinfo{author}{\bibfnamefont{T.~C.~H.} \bibnamefont{Liew}},
  \bibinfo{author}{\bibfnamefont{D.}~\bibnamefont{Comber-Todd}},
  \bibinfo{author}{\bibfnamefont{A.}~\bibnamefont{Nalitov}},
  \bibinfo{author}{\bibfnamefont{M.}~\bibnamefont{Steger}},
  \bibinfo{author}{\bibfnamefont{K.}~\bibnamefont{West}},
  \bibinfo{author}{\bibfnamefont{L.}~\bibnamefont{Pfeiffer}},
  \bibinfo{author}{\bibfnamefont{D.~W.} \bibnamefont{Snoke}},
  \bibnamefont{et~al.}, \bibinfo{journal}{Phys. Rev. Lett.}
  \textbf{\bibinfo{volume}{120}}, \bibinfo{pages}{065301}
  (\bibinfo{year}{2018}),
  \urlprefix\url{https://link.aps.org/doi/10.1103/PhysRevLett.120.065301}.

\bibitem[{\citenamefont{Voigt}(1902)}]{Voigt1902}
\bibinfo{author}{\bibfnamefont{W.}~\bibnamefont{Voigt}},
  \bibinfo{journal}{Philosophical Magazine Series}
  \textbf{\bibinfo{volume}{4}}, \bibinfo{pages}{90} (\bibinfo{year}{1902}).

\bibitem[{\citenamefont{Berry and Dennis}(2003)}]{Berry2003}
\bibinfo{author}{\bibfnamefont{M.~V.} \bibnamefont{Berry}} \bibnamefont{and}
  \bibinfo{author}{\bibfnamefont{M.~R.} \bibnamefont{Dennis}},
  \bibinfo{journal}{Proc. R. Soc. Lond. A} \textbf{\bibinfo{volume}{459}},
  \bibinfo{pages}{1261} (\bibinfo{year}{2003}).

\bibitem[{\citenamefont{Landau and Lifshitz}(1984)}]{Landau8}
\bibinfo{author}{\bibfnamefont{L.~D.} \bibnamefont{Landau}} \bibnamefont{and}
  \bibinfo{author}{\bibfnamefont{E.~M.} \bibnamefont{Lifshitz}},
  \emph{\bibinfo{title}{Electrodynamics of Continuous Media}}
  (\bibinfo{publisher}{Butterworth-Heinemann}, \bibinfo{year}{1984}).

\bibitem[{\citenamefont{Sturm et~al.}(2020)\citenamefont{Sturm, Zviagin, and
  Grundmann}}]{Sturm2020}
\bibinfo{author}{\bibfnamefont{C.}~\bibnamefont{Sturm}},
  \bibinfo{author}{\bibfnamefont{V.}~\bibnamefont{Zviagin}}, \bibnamefont{and}
  \bibinfo{author}{\bibfnamefont{M.}~\bibnamefont{Grundmann}},
  \bibinfo{journal}{Phys. Rev. Materials} \textbf{\bibinfo{volume}{4}},
  \bibinfo{pages}{055203} (\bibinfo{year}{2020}),
  \urlprefix\url{https://link.aps.org/doi/10.1103/PhysRevMaterials.4.055203}.

\bibitem[{\citenamefont{Richter et~al.}(2017)\citenamefont{Richter, Michalsky,
  Sturm, Rosenow, Grundmann, and Schmidt-Grund}}]{PhysRevA.95.023836}
\bibinfo{author}{\bibfnamefont{S.}~\bibnamefont{Richter}},
  \bibinfo{author}{\bibfnamefont{T.}~\bibnamefont{Michalsky}},
  \bibinfo{author}{\bibfnamefont{C.}~\bibnamefont{Sturm}},
  \bibinfo{author}{\bibfnamefont{B.}~\bibnamefont{Rosenow}},
  \bibinfo{author}{\bibfnamefont{M.}~\bibnamefont{Grundmann}},
  \bibnamefont{and}
  \bibinfo{author}{\bibfnamefont{R.}~\bibnamefont{Schmidt-Grund}},
  \bibinfo{journal}{Phys. Rev. A} \textbf{\bibinfo{volume}{95}},
  \bibinfo{pages}{023836} (\bibinfo{year}{2017}),
  \urlprefix\url{https://link.aps.org/doi/10.1103/PhysRevA.95.023836}.

\bibitem[{\citenamefont{Richter et~al.}(2019)\citenamefont{Richter, Zirnstein,
  Z\'u\~niga P\'erez, Kr\"uger, Deparis, Trefflich, Sturm, Rosenow, Grundmann,
  and Schmidt-Grund}}]{Richter2019}
\bibinfo{author}{\bibfnamefont{S.}~\bibnamefont{Richter}},
  \bibinfo{author}{\bibfnamefont{H.-G.} \bibnamefont{Zirnstein}},
  \bibinfo{author}{\bibfnamefont{J.}~\bibnamefont{Z\'u\~niga P\'erez}},
  \bibinfo{author}{\bibfnamefont{E.}~\bibnamefont{Kr\"uger}},
  \bibinfo{author}{\bibfnamefont{C.}~\bibnamefont{Deparis}},
  \bibinfo{author}{\bibfnamefont{L.}~\bibnamefont{Trefflich}},
  \bibinfo{author}{\bibfnamefont{C.}~\bibnamefont{Sturm}},
  \bibinfo{author}{\bibfnamefont{B.}~\bibnamefont{Rosenow}},
  \bibinfo{author}{\bibfnamefont{M.}~\bibnamefont{Grundmann}},
  \bibnamefont{and}
  \bibinfo{author}{\bibfnamefont{R.}~\bibnamefont{Schmidt-Grund}},
  \bibinfo{journal}{Phys. Rev. Lett.} \textbf{\bibinfo{volume}{123}},
  \bibinfo{pages}{227401} (\bibinfo{year}{2019}),
  \urlprefix\url{https://link.aps.org/doi/10.1103/PhysRevLett.123.227401}.

\bibitem[{\citenamefont{Hamilton}(1837)}]{Hamilton1837}
\bibinfo{author}{\bibfnamefont{W.~R.} \bibnamefont{Hamilton}},
  \bibinfo{journal}{Trans. Royal Irish Acad.} \textbf{\bibinfo{volume}{17}},
  \bibinfo{pages}{1} (\bibinfo{year}{1837}).

\bibitem[{\citenamefont{Lloyd}(1837)}]{Lloyd1837}
\bibinfo{author}{\bibfnamefont{H.}~\bibnamefont{Lloyd}},
  \bibinfo{journal}{Trans. Roayl Irish Acad.} \textbf{\bibinfo{volume}{17}},
  \bibinfo{pages}{145} (\bibinfo{year}{1837}).

\bibitem[{\citenamefont{Ter\ifmmode~\mbox{\c{c}}\else \c{c}\fi{}as
  et~al.}(2014)\citenamefont{Ter\ifmmode~\mbox{\c{c}}\else \c{c}\fi{}as,
  Flayac, Solnyshkov, and Malpuech}}]{Tercas2014}
\bibinfo{author}{\bibfnamefont{H.}~\bibnamefont{Ter\ifmmode~\mbox{\c{c}}\else
  \c{c}\fi{}as}}, \bibinfo{author}{\bibfnamefont{H.}~\bibnamefont{Flayac}},
  \bibinfo{author}{\bibfnamefont{D.~D.} \bibnamefont{Solnyshkov}},
  \bibnamefont{and} \bibinfo{author}{\bibfnamefont{G.}~\bibnamefont{Malpuech}},
  \bibinfo{journal}{Phys. Rev. Lett.} \textbf{\bibinfo{volume}{112}},
  \bibinfo{pages}{066402} (\bibinfo{year}{2014}),
  \urlprefix\url{https://link.aps.org/doi/10.1103/PhysRevLett.112.066402}.

\bibitem[{sup()}]{suppl}
\bibinfo{note}{See Supplemental Material at [URL will be inserted by
  publisher].}

\bibitem[{\citenamefont{Fieramosca et~al.}(2019)\citenamefont{Fieramosca,
  Polimeno, Lerario, De~Marco, De~Giorgi, Ballarini, Dominici, Ardizzone,
  Pugliese, Maiorano et~al.}}]{fieramosca2019chromodynamics}
\bibinfo{author}{\bibfnamefont{A.}~\bibnamefont{Fieramosca}},
  \bibinfo{author}{\bibfnamefont{L.}~\bibnamefont{Polimeno}},
  \bibinfo{author}{\bibfnamefont{G.}~\bibnamefont{Lerario}},
  \bibinfo{author}{\bibfnamefont{L.}~\bibnamefont{De~Marco}},
  \bibinfo{author}{\bibfnamefont{M.}~\bibnamefont{De~Giorgi}},
  \bibinfo{author}{\bibfnamefont{D.}~\bibnamefont{Ballarini}},
  \bibinfo{author}{\bibfnamefont{L.}~\bibnamefont{Dominici}},
  \bibinfo{author}{\bibfnamefont{V.}~\bibnamefont{Ardizzone}},
  \bibinfo{author}{\bibfnamefont{M.}~\bibnamefont{Pugliese}},
  \bibinfo{author}{\bibfnamefont{V.}~\bibnamefont{Maiorano}},
  \bibnamefont{et~al.}, \bibinfo{journal}{arXiv:1912.09684}
  (\bibinfo{year}{2019}).

\end{thebibliography}

\end{document}